\documentclass[useAMS,usenatbib,usegraphicx]{mn2e}

\title[Circumstellar disks in the $\sigma$\,Orionis cluster]{An L$'$-band survey for circumstellar disks around low-mass stars in the young $\sigma$\,Orionis cluster}
\author[Oliveira, Jeffries \& van Loon]{J.M. Oliveira\thanks{E-mail:
joana@astro.keele.ac.uk}, R.D. Jeffries and J.Th. van Loon\\
School of Chemistry and Physics, Keele University, Keele, Staffordshire
ST5 5BG, UK}
\begin{document}

\date{}

\pagerange{\pageref{firstpage}--\pageref{lastpage}} \pubyear{2003}

\maketitle

\label{firstpage}

\begin{abstract}
We present new K- and L$'$-band imaging of a representative sample of members of the young 3$-$5\,Myr old $\sigma$\,Orionis cluster. We identified objects with $(K-L')$ excess by analysing colour-colour diagrams and comparing the observations with empirical main-sequence colours. The derived disk frequency depends on the method used: (54$\pm$15)\% if measured directly from the $JHKL'$ colour-colour diagram; or (46$\pm$14)\% if excesses are computed with respect to predicted photospheric colours (according to the objects spectral types, 2-$\sigma$ excess detections). We compare the $(K-L')$ excess with other indicators and show that this is a robust and reliable disk indicator. We also compare the derived disk frequency with similarly aged clusters and discuss possible implications for disk lifetimes. The computed age of the $\sigma$\,Ori cluster is very important: a cluster age of 3\,Myr would support the overall disk lifetime of 6\,Myr proposed in the literature, while an age $>$\,4\,Myr would point to a slower disk destruction rate. 

\end{abstract}


\section{Introduction}
Disk-like structures are believed to be ubiquitous around young protostars. These disks are dissipated very early in pre-main-sequence (PMS) evolution, perhaps by powerful stellar jets/outflows or photodissociation by the far-ultraviolet flux from nearby massive OB stars. Despite their short lives, the timescales and mass dependence of disk dissipation have far reaching consequences in astrophysics: the efficiency of disk depletion could be the strongest factor in determining the timescales on which planets form in a particular stellar system \citep*{haisch01}, or whether they form at all \citep{brandner00}. Disks probably play a significant role in early angular momentum regulation and the dissipation timescale is thought to control the spread in rotation rates of young stars \citep*{sills00}. Stars may accrete a significant fraction of their final mass from a circumstellar disk, so the timescale and mass dependence of that accretion influences PMS evolution and thus attempts to estimate ages and masses from evolutionary PMS models \citep{comeron03}. The mass dependence of disk frequencies can provide a stern test for low-mass stellar and brown dwarf formation theories. For instance, models involving competitive accretion and subsequent ejection of brown dwarfs from protostellar aggregates \citep{reipurth00,bate03} may imply shorter disk dissipation times for the lower mass fragments.

Observed disk frequencies in samples of young stars with different ages, masses
and environments provide an empirical determination of disk lifetimes. Judging from L-band excesses, young clusters exhibit high disk frequencies ($\ga$\,80\%, e.g.\ the Trapezium cluster: \citealt{lada00}) up to ages of $\sim$\,1.5\,Myr, which then decrease rapidly with age: at $\sim$\,3\,Myr, 50\% of disks have been dissipated, and the timescale for all cluster members to lose their disks may be as short as $\sim$\,6\,Myr \citep{haisch01}. Such timescales have been questioned by a high disk frequency in the 9\,Myr $\eta$\,Chamaeleontis cluster, a sparsely populated cluster with no massive stars \citep*{lyo03}.

$\sigma$\,Orionis is a Trapezium-like system with an O9.5\,V primary. The population of low-mass stars spatially clustered around this system was discovered as bright X-ray sources in ROSAT images, and follow-up optical spectroscopy confirmed most sources as PMS stars \citep{wolk96,walter97}. This association is young, nearby and affected by low reddening, making it an ideal target to analyse the PMS population even down to brown dwarfs \citep[e.g.\,][]{bejar01, barrado03,kenyon03} and isolated planetary mass objects \citep{osorio00}. Furthermore, at an age of 3$-$5\,Myr \citep[e.g.\,][]{oliveira02,osorio02,jayawardhana03}, the $\sigma$\,Orionis cluster is at a crucial stage in terms of disk evolution and it is therefore a key case to better constrain disk dissipation timescales. Recently, a possible proto-planetary disk, apparently on the process of being dissipated, has been discovered very close to $\sigma$\,Ori \citep{loon03}.

The $K_{\rm s}$-excess disk frequency is 5$-12$\,\% for the low-mass and brown dwarf members of the $\sigma$\,Orionis cluster \citep{oliveira02,barrado03}. On the other hand, the presence of strong H$\alpha$ emission suggests accretion disk frequencies as high as 30\% \citep{osorio02}. However, the most reliable method to determine the disk frequency in a low-mass population is by measuring the $(K-L)$ colours and deriving colour excesses \citep[e.g.\,][]{wood02}. \citet{jayawardhana03} have obtained L-band observations of 6 $\sigma$\,Ori cluster members, finding two with a $(K-L)$ excess. The significance of this result is obviously limited by the size of the sample. We have performed L$'$-band (3.8\,$\mu$m) observations of a representative sample of 28 cluster members, using the newly installed imager UIST at the United Kingdom Infrared Telescope (UKIRT). Young stars are well known for their variability across the spectrum including infrared (IR) wavelengths \citep{carpenteretal01,carpenter02}, therefore we have obtained nearly simultaneous K-band observations for all our targets. In this paper, we describe the results of this survey, and discuss our derived disk frequency within the framework of disk destruction timescales by comparing with similar surveys in other young clusters.

\section{Cluster members and properties}

\subsection{Sample of cluster members}

We have an on-going program to observe in the L-band $\sigma$\,Ori cluster members identified at optical wavelengths. We describe here the observations of 28 of the brightest cluster members: their positions, $I_{\rm c}$ magnitudes, 2MASS (Two Micron All Sky Survey) $J, H$ and $K_{\rm s}$ magnitudes, the new K- and L$'$-band  magnitudes and identifications are listed in Table\,\ref{obs_table}. Some sources were first identified as ROSAT X-ray sources and photometric cluster candidates by \citet[][ W96]{wolk96} while other objects are photometric candidates identified by \citet[][ B01]{bejar01}. \citet[][ ZO02]{osorio02} have spectroscopically confirmed cluster membership for both these sets of objects. The remaining objects are spectroscopic cluster members found by \citet[][ K03]{kenyon03}. The I-band magnitudes are from either \citet{bejar01} or \citet{kenyon03}. The brighter objects in the sample are mostly from the X-ray selected sample \citep{wolk96} while the fainter objects were photometrically and spectroscopically selected. In Sect.\,5.1 we discuss the effects of possible selection biases on our results.

Searching for circumstellar disks around these 24 objects is the main goal of these observations. To this sample we have added 4 objects that are known IRAS sources in the region (no reference entry in Table\,\ref{obs_table}). They have been confirmed by \citet{oliveira03b} (see also \citealt{oliveira03a}) as mid-infrared sources with spectral energy distributions (SEDs) consistent with them being young stars with dusty circumstellar disks; thus, based on their location, youth and infrared excesses they are also likely to be members of the $\sigma$\,Ori cluster. 

\begin{table*}
\centering
\begin{minipage}{180mm}
\caption{Cluster members imaged with UIST at UKIRT. Column 1 is the target number, columns 2 and 3 are the targets' 2MASS positions, column 4 is the $I_{\rm c}$ magnitude from the literature (not available for the IRAS sources), columns 5$-$7 are the 2MASS $J, H$ and $K_{\rm s}$ magnitudes (typical uncertainties of about 0.03\,mag), columns 8$-$11 are the measured $K$ and $L'$ magnitudes in the MKO system with their uncertainties and columns 12 and 13 give the target identification and  references (see text). Target identifications of the type J053920.5$-$022737 are an abbreviation of the IAU denomination SOri\,J053920.5$-$022737.}
\label{obs_table}
\begin{tabular}{rcclrrrrlrlll}
\hline
  & \multicolumn{2}{c}{J2000 Position} && \multicolumn{3}{c}{2MASS magnitudes} & \multicolumn{4}{c}{MKO magnitudes}&\multicolumn{2}{c}{Identification}\\
  &    ra     &     dec  &  \multicolumn{1}{c}{$I_{\rm c}$} & \multicolumn{1}{c}{$J$}   &   \multicolumn{1}{c}{$H$}  &\multicolumn{1}{c}{K$_{\rm s}$}&\multicolumn{2}{c}{$K$} & \multicolumn{2}{c}{$L'$}& \multicolumn{1}{l}{name} & \multicolumn{1}{l}{reference}\\
  & ($^{h\,\,m\,\,s}$)&($^{d\,\,m\,\,s}$) & (mag) &(mag) & (mag) & (mag) & \multicolumn{2}{c}{(mag)}&\multicolumn{2}{c}{(mag)}& \\
\hline
 1& 5 38 33.68       & $-$2 44 14.2&      & 10.13&  9.28&  8.66&  8.600& 0.00\rlap{2}&  7.71& 0.02& TX\,Ori&\\
 2& 5 38 48.04       & $-$2 27 14.2&12.08& 10.16&  9.46&  9.19&  9.208& 0.00\rlap{4}&  8.65& 0.08& 4771-899&W96, ZO02\\
 3& 5 38 27.26       & $-$2 45 09.7&12.82& 11.96& 10.79&  9.94&  9.729&
 0.00\rlap{6}&  8.64& 0.04& 4771-41, V505\,Ori&W96, ZO02\\
 4& 5 40 08.89       & $-$2 33 33.7&      & 11.50& 10.55&  9.91&  9.812& 0.00\rlap{6}&  8.76& 0.06& Haro\,5-39&\\
 5& 5 39 39.83       & $-$2 33 16.0&      & 12.22& 10.96& 10.07& \llap{1}0.387& 0.00\rlap{9}&  9.09& 0.06& V\,603\,Ori&\\
 6& 5 39 39.82       & $-$2 31 21.8&      & 11.84& 10.90& 10.22&  9.831& 0.00\rlap{6}&  8.48& 0.07& V\,510\,Ori&\\
 7& 5 38 38.23       & $-$2 36 38.4&12.37& 11.16& 10.46& 10.31& \llap{1}0.232& 0.00\rlap{8}& \llap{1}0.18& 0.03& r053838$-$0236&W96,ZO02\\
 8& 5 38 31.58       & $-$2 35 14.9&13.52& 11.52& 10.70& 10.35& \llap{1}0.565& 0.00\rlap{9}&  9.63& 0.05& r053831$-$0235&W96,ZO02\\
 9& 5 38 40.27       & $-$2 30 18.5&12.80& 11.51& 10.76& 10.40& \llap{1}0.350& 0.00\rlap{8}&  9.75& 0.07& r053840$-$0230&W96,ZO02\\
\llap{1}0& 5 38 49.17& $-$2 38 22.2&12.88& 11.39& 10.66& \llap{1}0.51& 10.456& 0.009& \llap{1}0.22& 0.04& r053849$-$0238&W96,ZO02\\
\llap{1}1& 5 39 05.41& $-$2 32 30.3&12.66& 11.55& 10.86& \llap{1}0.67& 10.618& 0.009& \llap{1}0.44& 0.05& 4771-1075&W96,ZO02\\
\llap{1}2& 5 39 11.63& $-$2 36 02.9&12.78& 11.62& 10.97& \llap{1}0.75& 10.72 & 0.01 & \llap{1}0.53& 0.09& 4771-1038&W96,ZO02\\
\llap{1}3& 5 39 20.44& $-$2 27 36.8&13.51& 12.15& 11.42& \llap{1}1.17& 11.12 & 0.01 & \llap{1}0.91& 0.06& J053920.5$-$022737&B01,ZO02\\
\llap{1}4& 5 40 01.96& $-$2 21 32.6&14.32& 12.34& 11.58& \llap{1}1.25& 11.18 & 0.01 & \llap{1}0.67& 0.05& J054001.8$-$022133&B01,ZO02\\
\llap{1}5& 5 38 47.55& $-$2 27 12.0&14.46& 12.14& 11.50& \llap{1}1.27& 11.20 & 0.01 & \llap{1}0.83& 0.05& J053847.5$-$022711&B01,ZO02\\
\llap{1}6& 5 38 20.21& $-$2 38 01.6&14.41& 12.58& 11.86& \llap{1}1.61& 11.51 & 0.02 & \llap{1}1.16& 0.11& J053820.1$-$023802&B01,ZO02\\ 
\llap{1}7& 5 39 51.73& $-$2 22 47.2&14.59& 12.60& 12.01& \llap{1}1.68& 11.65 & 0.02 & \llap{1}1.02& 0.04& J053951.6$-$022248&B01,ZO02\\
\llap{1}8& 5 39 40.98& $-$2 16 24.4&15.84& 12.87& 12.15& \llap{1}1.74& 11.72 & 0.02 & \llap{1}1.00& 0.05& J053941.0$-$021624&K03\\
\llap{1}9& 5 38 27.51& $-$2 35 04.2&14.50& 12.83& 12.11& \llap{1}1.86& 11.72 & 0.02 & \llap{1}1.18& 0.05& J053827.4$-$023504&B01,ZO02\\
\llap{2}0& 5 39 58.26& $-$2 26 18.8&14.19& 12.78& 12.05& \llap{1}1.83& 11.77 & 0.02 & \llap{1}1.60& 0.05& J053958.1$-$022619&B01,ZO02\\ 
\llap{2}1& 5 39 22.87& $-$2 33 33.1&14.16& 12.83& 12.13& \llap{1}1.87& 11.82 & 0.02 & \llap{1}1.55& 0.04& r053923$-$0233&W96,ZO02\\
\llap{2}2& 5 39 07.59& $-$2 28 23.4&14.33& 12.88& 12.14& \llap{1}1.96& 12.12 & 0.03 & \llap{1}1.61& 0.06& r053907$-$0228&W96,ZO02\\
\llap{2}3& 5 40 09.33& $-$2 25 06.7&15.11& 13.15& 12.50& \llap{1}2.15& 12.06 & 0.03 & \llap{1}1.55& 0.04& J054009.3$-$022507&K03\\
\llap{2}4& 5 40 01.01& $-$2 19 59.8&15.02& 13.10& 12.50& \llap{1}2.25& 12.18 & 0.02 & \llap{1}1.90& 0.03& J054551.0$-$021960&K03\\
\llap{2}5& 5 39 14.47& $-$2 28 33.4&14.75& 13.34& 12.65& \llap{1}2.34& 12.32 & 0.03 & \llap{1}1.79& 0.10& J053914.5$-$022834&B01,ZO02\\
\llap{2}6& 5 38 17.47& $-$2 09 23.6&14.94& 13.28& 12.66& \llap{1}2.36& 12.46 & 0.03 & \llap{1}2.14& 0.04& J053817.5$-$020924&K03\\
\llap{2}7& 5 37 54.53& $-$2 58 26.5&15.51& 13.31& 12.71& \llap{1}2.41& 12.29 & 0.03 & \llap{1}1.70& 0.10& J053754.5$-$025827&K03\\
\llap{2}8& 5 37 58.40& $-$2 41 26.2&15.36& 13.29& 12.70& \llap{1}2.42& 12.39 & 0.03 & \llap{1}1.62& 0.11& J053758.4$-$024126&K03\\
\hline
\end{tabular}
\end{minipage}
\end{table*}

\subsection{Cluster age}

\begin{figure}
\includegraphics[totalheight=8.5cm]{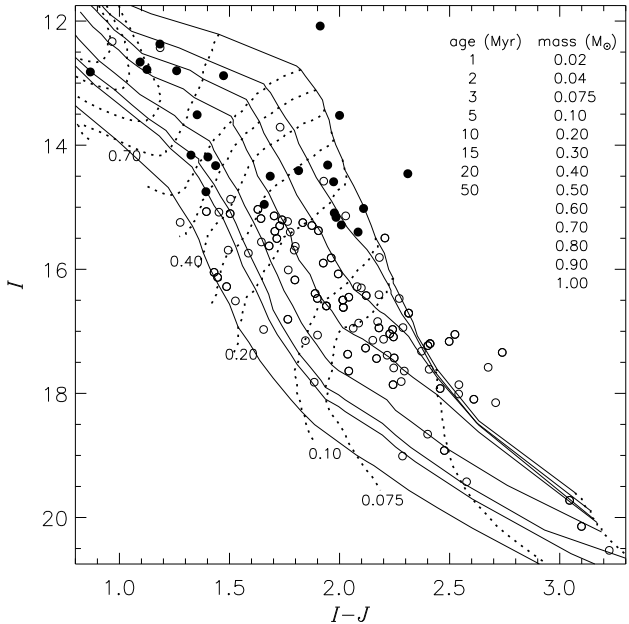}
\caption{$I/(I-J)$ colour-magnitude diagram of the $\sigma$\,Ori cluster. All the stars in this diagram have been spectroscopically confirmed as cluster members \citep{osorio02,barrado03,kenyon03}. The filled symbols are the objects we have observed in the K- and L$'$-band. We overplotted evolutionary tracks (dotted lines) and isochrones (solid lines) from \citet[][ see text]{baraffe98}. }
\label{ij_cmd}
\end{figure} 

The age determination for clusters at very young ages is always an uncertain affair. In OB associations, several methods can be used: the properties of the massive O-stars, low-mass stellar isochrone fitting, and lithium abundance evolution in the association.

The age of the multiple system $\sigma$\,Ori is estimated to be 1.7$-$7\,Myr based on stellar properties and its membership of the Orion\,OB1b group \citep*[][ and references therein]{brown94}. Using isochrones from several authors, \citet*[][ and references therein]{bejar99} determined a low-mass star cluster age between 1$-$5\,Myr; using a different sample of low-mass cluster members, \citet{oliveira02} determined an isochronal median age of 4.2$^{+2.7}_{-1.5}$\,Myr, where the quoted errors are due to the uncertainties of the Hipparcos distance to $\sigma$\,Ori. \citet{osorio02} compared the lithium abundance in low-mass cluster members with theoretical predictions from several authors. They found no evidence of appreciable lithium destruction and  they inferred an upper limit to the cluster age of 8\,Myr.

In Fig.\,\ref{ij_cmd} we plot the $I/(I-J)$ colour-magnitude diagram for the present sample. We use the \citet{baraffe98} evolutionary models with the mixing length parameter set to 1.0 pressure scale height for lower mass objects and to 1.9 for $M > 0.62$\,M$_{\sun}$. We have opted to use these models because they incorporate model atmospheres and predict PMS magnitudes without the use of empirical effective temperature-colour relations. We have computed the age and mass of each object; two objects (from the 24 with I-band magnitudes) fall outside the Baraffe grid of models --- they appear to be very young (age\,$<$\,1\,Myr). The objects have masses in the range 1.0$-$0.13\,M$_{\sun}$. The sample has a median age of 3.6\,Myr (at the Hipparcos distance of 352\,pc), consistent with previous age determinations (see above). The conservatively large uncertainty in the Hipparcos distance (352$^{+166}_{-85}$\,pc) produces median age uncertainties as computed by \citet{oliveira02}.

The most obvious thing from this colour-magnitude diagram is the large apparent age spread ($<$\,1 to 20\,Myr). \citet{hartmann01} has analysed many possible sources of uncertainty in the determined stellar ages in star forming regions. He concluded that observational errors probably account for a large fraction of the observed age spread in such regions. Therefore, it is still not firmly established whether observed age spreads are real effects. 

The main sources of uncertainty in the case of the present data set are photometric variability and unresolved binaries. Unresolved binaries would tend to make objects appear brighter and redder, making them look younger than isolated cluster siblings. This is one reason why we consider the true age of the cluster likely to be older than the median age of the cluster (3.6\,Myr, see above). Another effect, seldom considered, is described by \citet{comeron03}: they propose that underluminous cluster members in the Lupus\,3 dark cloud provide evidence for accretion-modified evolution. These objects, if not {\it a priori} excluded from a cluster sample, would appear older than the rest of the cluster members.

The effect that we consider to be the major cause of the large age spread we observe in the $\sigma$\,Ori cluster is variability. We make use of 2MASS J-band magnitudes and I-band magnitudes from the literature. PMS stars are known to be variable in $J, H$ and $K$ \citep{carpenteretal01,carpenter01} and we found evidence of K-band variability in this sample (Sect.\,3.2). They found a dispersion among variable stars of about 0.1\,mag in $J$; if the effect in the I-band is of similar amplitude and the variability in the two bands is independent, an uncertainty in the $(I-J)$ colour of more than 0.1\,mag would not be surprising. If variability is related to the presence of a circumstellar disk, then the uncertainty could be even larger \citep{carpenteretal01,carpenter01}. Indeed, from the 6 objects we found to be variable in the K-band (Sect.\,3.2), two objects have no I-band measurements; one object falls outside the model grid i.e.\ appears too young; of the remaining 3 objects,  2 have computed ages of 16\,Myr and 20\,Myr. Although limited by small number statistics, 3 out of 4 K-band variable objects have ``anomalous'' ages, compatible with the spread being caused by variability.

Therefore, we believe that most of the spread we observe in Fig.\,\ref{ij_cmd} is not a real age spread, and that the likely age of the cluster is larger than 3.6\,Myr with an upper limit of 8\,Myr imposed by \citet{osorio02}. We will take into account the age uncertainty in our subsequent analysis. This short summary of the problems in determining the age of young clusters highlights the importance of using consistent age determinations, for instance when trying to infer an overall disk destruction timescale (Sect.\,5.3).

\subsection{Reddening}

\begin{figure}
\includegraphics[totalheight=8.5cm]{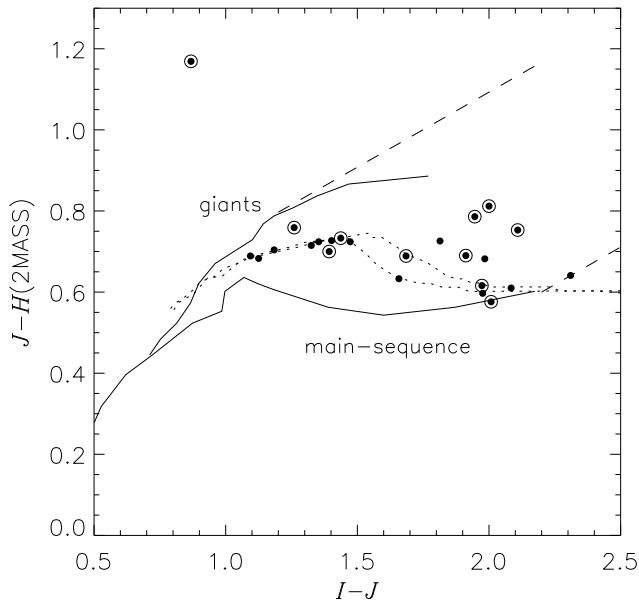}
\caption{$IJH$ colour-colour diagram for the target sample. The solid lines are the empirical loci for main-sequence and giant stars (to spectral type M5) and the dashed lines are reddening band \citep[from][ converted to the appropriate photometric systems]{bessell88}. Isochrones for 3\,Myr (upper dotted line) and 5\,Myr (lower dotted line) from \citet{baraffe98} are also plotted. The double-circle symbols are objects that are found to exhibit $(K-L')$ excess in the $IJKL'$ colour-colour diagram (see Sect.\,4.1). The object with $(J-H)\sim 1.2$ is a bright cluster member and it is the only object with a large $(H-K_{\rm s})$ excess (Fig.\,\ref{colour_colour}). This diagram shows no evidence for significant reddening towards these cluster members.}
\label{reddening}
\end{figure} 

The average reddening towards the O-star $\sigma$\,Ori is quite low: $E(B-V)=0.05$ \citep[e.g.\,][]{brown94}. However, a superficial analysis (see below) of traditional colour-colour diagrams seems to hint at some amount of reddening towards cluster members. Therefore, we have decided to investigate the question of reddening for the target sample.

Fig.\,\ref{reddening} shows the $IJH$ colour-colour diagram for the target sample. This diagram has been used with success to compute individual reddening to young stars in clusters \citep{lucas00,thompson03}. The solid lines are the empirical loci for main-sequence and giant stars (from \citet{bessell88} converted to 2MASS magnitudes using the transformations from \citet{carpenter02}) and the dashed lines are the reddening bands \citep{rieke85}. Traditionally, one attempts to de-redden individual objects with respect to main-sequence colours. But PMS stars have lower gravities than field dwarfs and thus populate the region between the main-sequence and giant loci. The dotted lines are isochrones from \citet{baraffe98} for 3\,Myr and 5\,Myr; the target colours in the diagram are quite well represented by the isochrones. According to I.\, Baraffe (private communication) we should be cautious when using H-band magnitudes produced by their models because of possible shortcomings in the water line-list used (the H-band includes several strong water absorption bands). Nevertheless, very few objects show any evidence of significant reddening. The few objects that seem reddened actually are found to exhibit a $(K-L')$ excess that indicates the presence of a circumstellar disk (see Sect.\,4). We could use the reddening measured towards $\sigma$\,Ori, but it is so small that its effect is negligible --- especially at IR wavelengths. Therefore, the results described in this paper were obtained without de-reddening the target colours and magnitudes. 

\section{Infrared Observations}

\subsection{K and L$'$-band imaging}

In order to compute meaningful $(K-L')$ excesses, it is essential that we are able to detect the stellar photospheres both in the K- and L$'$-bands. We used the 2MASS $K_{\rm s}$ magnitudes of the targets to estimate exposure times in the K-band and use the \citet{baraffe98} evolutionary models (see previous section) to estimate L'-band magnitudes. Every target was detected in both bands. The observations in the two filters were consecutive, in order to minimise the effect of variability in the measured $(K-L')$ colours.

The observations were performed at the UKIRT on the 17th and 19th January 2003, with the newly installed imager-spectrometer UIST. We observed with the broad-band K and L$'$ filters on the Mauna Kea Observatory Near-Infrared (MKO-NIR) system \citep*{tokunaga02}. For both K- and L$'$-band observations, we used the 0.06$\arcsec$ pixel scale and windowed the detector to 512\,$\times$\,512 pixels to reduce overheads, resulting in a field of view of 31$\arcsec \times 31\arcsec$. Typically this results in no more than one cluster member per field, so that exposure times are tailored for each target. Adverse weather conditions meant that we only could observe for about 30\% of the observing time allocated to this project; from our original target list we observed the 28 brightest (in $K_{\rm s}$) targets. On the first night conditions were stable and dry, with typical seeing of the order of 0.7\,arcsecond (measured as the full-width at half-maximum of the K-band images); in the second night conditions were less stable, with seeing of about 1 arcsecond.

In the L$'$-band, on-target individual exposure times ranged from 1.5\,min to 15\,min; the brightest targets were observed on a 4-point dither pattern and the fainter targets on an 8-point dither pattern. At each dither position, 27 (sky-limited) exposures of 0.81\,s were co-added. An 8-point dither pattern is complete in approximately 3\,min (on-target) and the cycle was repeated until the requested exposure time was reached for each target. The frames were reduced and combined using ORAC-DR (the UKIRT data reduction and high level instrument control software), with dedicated recipes for the reduction of thermal imaging data. Sky background was removed by subtracting a median-average of the closest-in-time exposures at the different dither positions; flat-field frames were created by combining normalised object frames using the median at each pixel. Aperture photometry was performed on the mosaiced frames using the Starlink package GAIA (Graphical Astronomy and Image Analysis Tool), with apertures of 1.2$\arcsec$ and 1.8$\arcsec$ for the first and second nights respectively. The statistical signal-to-noise ratio of the L$'$-band photometry ranges from 50 to 9. To calibrate the observations we used the following standards \citep[see][]{leggett03}: HD\,40335 ($L'=6.45$\,mag) and SAO\,112626 ($L'=8.56$\,mag), respectively for the 4-point and 8-point dither patterns. Standards were observed every one to two hours.

In the K-band the targets are rather bright (Table\,\ref{obs_table}) so the exposure time was set to 10\,s. These observations were performed in a 5-point dither pattern. The K-band images were also reduced and analysed within ORAC-DR and GAIA. These observations were calibrated using the following standards: FS\,11 ($K=11.252$\,mag) and FS\,121 ($K=11.302$\,mag) from the list of UKIRT Faint Standards. The signal-to-noise ratio of the K-band photometry exceeds 30 and is more precise than the corresponding $L'$ measurement in all cases.

\subsection{K-band variability}

For the targets in Table\,\ref{obs_table} we have two independent measurements of K-band magnitude: $K_{\rm s}$ from the 2MASS All-Sky Catalog of Point Sources \citep{cutri03} --- data taken in October 1998 --- and the $K$ from the present observations. To convert the 2MASS photometry to the MKO system we first converted from 2MASS to the old UKIRT system using the transformation in \citet{carpenter01} and then from this system to the new MKO system using the transformation provided in \citet{hawarden01}.

\begin{figure}
\includegraphics[totalheight=7.8cm]{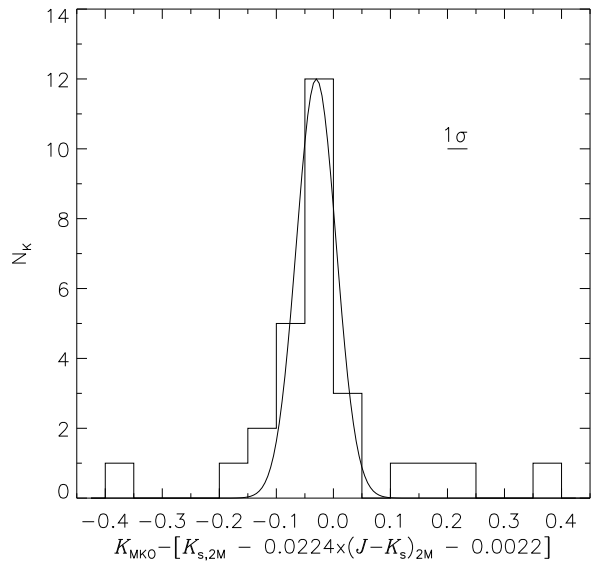}
\caption{Histogram of the difference between MKO $K$ and 2MASS $K_{\rm s}$ measurements. The 2MASS magnitudes were converted to the MKO photometric system (see text). The typical 1-$\sigma$ error-bar for these differences is indicated; the gaussian error distribution highlights how many objects are likely to be variable. There is a systematic shift of about 0.04\,mag between the 2 sets of measurements. Most objects do not show evidence of significant variability, with only 3 objects varying by more than 0.2\,mag.}
\label{k_band}
\end{figure} 

Fig.\,\ref{k_band} shows the histogram of the difference between the two sets of measurements in the MKO system. We can see that there is a systematic effect in the $K$ magnitude of the order of 0.04\,mag, that is acceptable within the measurement uncertainties --- it is not possible to say which of the steps of the transformation (2MASS to UKIRT or UKIRT to MKO) introduces this effect. Small differences can be easily explained by uncertainties in the calibrations and photometric system transformations. Nevertheless some objects show differences that clearly hint at K-band variability, namely targets 5, 6 and 8 and more marginally targets 3, 22 and 26. PMS stars are known to be variable at near-infrared wavelengths. \citet{carpenteretal01,carpenter02} found that $\sim$\,30\% of objects in Orion A and Chamaeleon I molecular clouds are variable. For more than 70\% of variable PMS stars, variability is characterised by small-amplitude ($\sim$0.1\,mag), essentially colourless, fluctuations that can be attributed to rotational modulation by cool spots. However, when the observed fluctuations are of larger amplitude and occur both in magnitudes and colours, then a combination of effects related with the presence of circumstellar disks is a likely cause: hot spots, variable circumstellar extinction and varying disk properties (accretion rate and/or inner disk temperature). Objects 5, 6 and 8 could fall in this latter category (they also have large $(K-L')$ excesses) and objects 3, 22 and 26 in the former but our data really does not allow us to say much more.

\section{Analysis of the $(K-L')$ colours of the cluster members}

\subsection{Colour-colour diagrams}

The most immediate method to search for circumstellar disks is to look for stars with $(H-K)$ or $(K-L)$ excesses in the $JHK$ or $JHKL$ colour-colour diagrams, that would indicate the presence of circumstellar dust material. Near-IR colours trace the warmer dust and the magnitude of a near-infrared excess depends critically on the position and temperature of the inner disk boundary. Therefore, disk frequencies determined from K-band excesses tend to be lower limits to the true frequency. At longer wavelengths, IR excesses grow rapidly and disk properties interfere less with the ability to detect disks. In particular for the L-band, disks remain optically thick over a wide range of disk masses, resulting in measurable $(K-L)$ excess down to very low disk masses \citep{wood02}.  Therefore, a $(K-L)$ excess is a very reliable disk diagnostic \citep{haisch01,wood02}. 

\begin{figure*}
\includegraphics[totalheight=17cm]{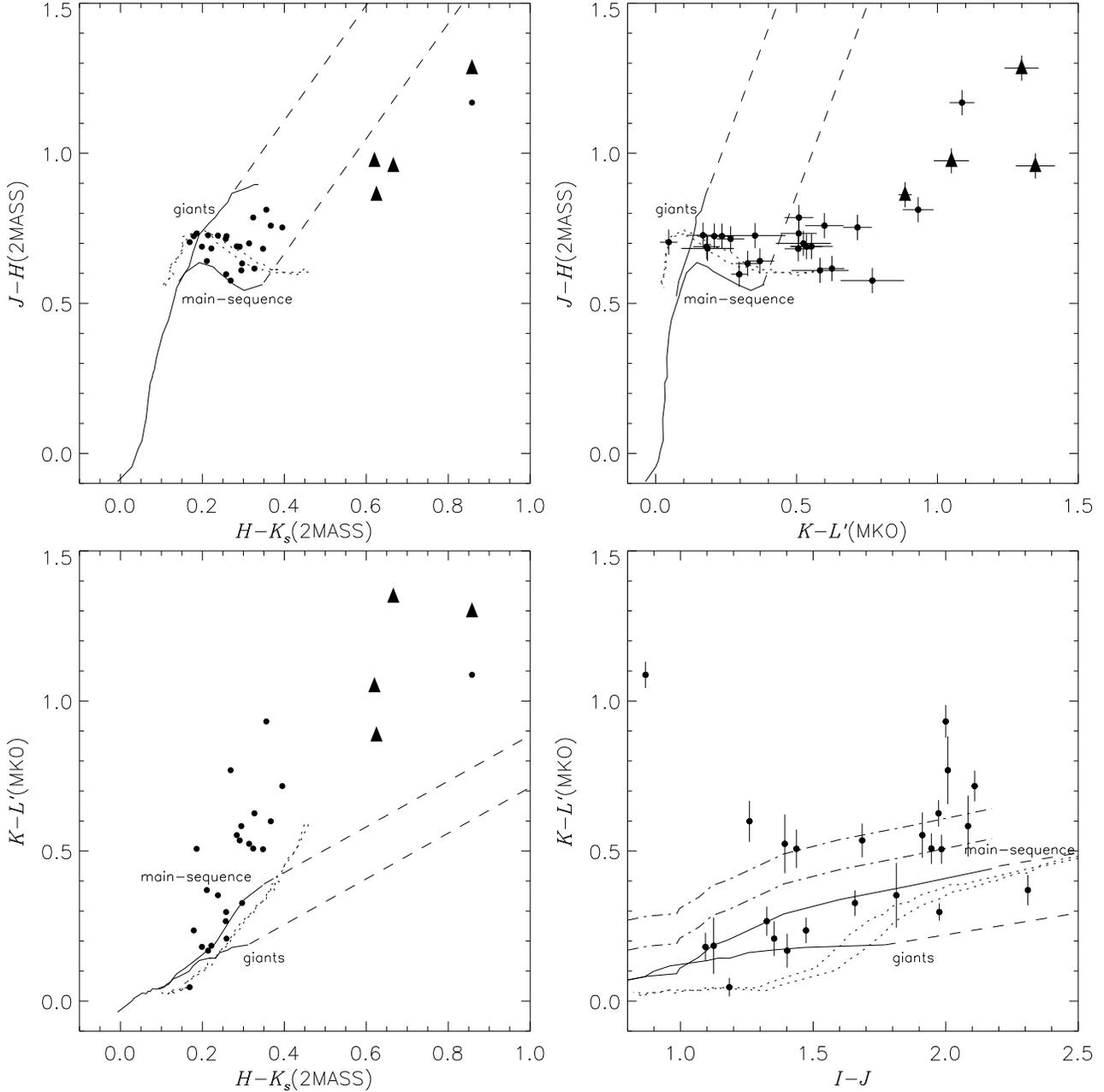}
\caption{Colour-colour diagrams for $\sigma$\,Ori cluster members; from left to right and top to bottom: $JHK_{\rm s}$ from 2MASS, $JHKL'$, $HK_{\rm s}KL'$ and $IJKL'$ from 2MASS and our UKIRT observations. The solid lines are the empirical loci for main-sequence (to spectral type M5) and giant stars as indicated by the labels and the dashed lines are reddening bands (see text). The dotted lines are \citet{baraffe98} isochrones for 3\,Myr and 5\,Myr. They illustrate the regions in the diagrams where young diskless PMS stars are expected to be: between the giant and the main-sequence loci. The triangle symbols represent IRAS sources, previously known to possess circumstellar disks. The dash-dotted lines in the $IJKL'$ diagram indicate the location of stars with excesses of 0.1\,mag and 0.2\,mag with respect to main-sequence colours.}
\label{colour_colour}
\end{figure*} 

Fig.\,\ref{colour_colour} shows several colour-colour diagrams for the 28 cluster members. As a clarification, throughout this paper $(J-H)$ and $(H-K_{\rm s})$ colours are always from 2MASS, while $(K-L')$ is always from our UKIRT observations. The IRAS sources are known to possess disks so they will be excluded from the discussion on disk frequencies (so as not to introduce any bias), but they are shown here to illustrate the position in the diagrams of classical T Tauri stars. The diagrams show the position of the objects with respect to the main-sequence and giant star loci and their reddening bands (see Fig.\,\ref{reddening}), converted to the 2MASS and MKO systems using transformations from \citet{bessell88}, \citet{hawarden01} and \citet{carpenter01}. When analysing such diagrams, stars are considered to exhibit an excess when they appear to the right of the reddening vector for late-type stars. From the $JHK_{\rm s}$ diagram (top left) it can be seen that (with the exception of the IRAS sources) only one object exhibits a considerable K-band excess. This scenario changes dramatically when looking at the $JHKL'$ diagram (top right): maybe as many as 13/24 objects (or 54\%) seem to have an L$'$-band excess. Simply counting stars with excess in such diagrams may still underestimate the true disk frequency: the reddening bands are rather broad, meaning that early-type stars with an excess might not be distinguishable from later-type stars with small reddening and no excess. 

The two diagrams at the bottom of Fig.\,\ref{colour_colour} largely remove this ambiguity. In the $HK_{\rm s}KL'$ and $IJKL'$ diagrams, the reddening bands are narrow and almost parallel to the empirical loci, so that stars with an excess in the L-band are clearly above these bands. The \citet{baraffe98} isochrones for 3\,Myr and 5\,Myr predict which region of the diagrams would be populated by diskless PMS stars.

The $IJKL'$ diagram (bottom right) is particularly clear, with two populations of stars: one with colours in between the giant and main-sequence loci and one of objects with colours more than 0.1\,mag above the empirical main-sequence locus. A legitimate question in the light of our concern about variability, is whether we can make use of $(I-J)$ colours that mix magnitudes obtained at different epochs ($J$ magnitudes are from 2MASS and $I$ magnitudes are from sources in the literature). The strength of this diagram, however, is that variability in $(I-J)$ would displace objects horizontally across the diagram (the same way as reddening) while an excess moves the objects almost vertically, so these effects act orthogonally. We can see from this diagram that, with respect to empirical main-sequence colours, 11/24 stars show no excess, 13/24 stars (or 54\%) show excesses $\ga 0.1$\,mag and 8/24 stars (or 33\%) show excesses $\ga 0.2$\,mag. 

\subsection{$(K-L')$ excesses}

\begin{table}
\centering
\begin{minipage}{86mm}
\caption{$(H-K_{\rm s})$ and $(K-L')$ excesses for the target sample. Column 2 gives the spectral types from the literature or, if none is available, estimated from photometric colours (see text). Columns 3 and 6 are the measured $(H-K_{\rm s})$ and $(K-L')$ colours; columns 4$-$5 and 7$-$8 are excesses with respect to photospheric colours and respective uncertainties. Column 9 is the measured EW[H$\alpha$] (in \AA). Column 10 provides the references for the spectral types and/or EW[H$\alpha$] (ZO02; \citealt{kholopov98}, Kh98).}
\label{excess_table}
\begin{tabular}{llllllllll}
\hline
  & spT  & \multicolumn{3}{c}{$(H-K_{\rm s})$}   & \multicolumn{3}{c}{$(K-L')$}  & H$\alpha$  &ref.\\
  &      &   \llap{o}bs.     & exc.    & \llap{er}ror            & obs. & exc.     &\llap{er}ror              & (\AA)          &\\
\hline
 1&  \llap{K}4  & \llap{0}.63   &0.49	&\llap{0}.05	     & 0.86    &  0.80 & \llap{0}.04 		&		 &\llap{K}h98\\
 2& \llap{K}7  & \llap{0}.28   &0.13    &\llap{0}.05	    & 0.55    &  0.44 & \llap{0}.08	       &3.1		&\llap{Z}O02\\
 3& \llap{K}7  & \llap{0}.86   &0.70    & \llap{0}.05	    & 1.09    &  0.98 & \llap{0}.05	      &\llap{5}3.5     &\llap{Z}O02\\
 4&  --  &\llap{0}.62  &\llap{$>$}0.22& &1.05 &\llap{$>$}0.59& & &     \\
 5&  --  &\llap{0}.86  &\llap{$>$}0.46& & 1.30&\llap{$>$}0.84& & &	\\
 6& \llap{K}8  & \llap{0}.67   &0.49 &\llap{0}.05	 & 1.35    &  1.22 & \llap{0}.07		       &		&\llap{K}h98\\
 7& \llap{K}8  & \llap{0}.17  &\llap{$-$}0.01 &\llap{0}.05 & 0.05    &\llap{$-$}0.08 &\llap{0}.04	      &2.9	       &\llap{Z}O02\\
 8& \llap{M}0  & \llap{0}.36  & 0.16    &\llap{0}.05	& 0.93    &  0.78 & \llap{0}.06 		      &4.5	       &\llap{Z}O02\\
 9& \llap{M}0  & \llap{0}.37  & 0.18    &\llap{0}.05	& 0.60    &  0.45 & \llap{0}.07 		      &6.7	       &\llap{Z}O02\\
\llap{1}0& \llap{M}0.5& \llap{0}.18  &\llap{$-$}0.03 &\llap{0}.05 & 0.24    &  0.07 & \llap{0}.05		      &2.6	       &\llap{Z}O02\\
\llap{1}1& \llap{K}7  & \llap{0}.20  & 0.04     &\llap{0}.05	& 0.18    &  0.07 & \llap{0}.05 		      &0.7	       &\llap{Z}O02\\
\llap{1}2& \llap{K}8  & \llap{0}.22  & 0.05     &\llap{0}.05	& 0.18    &  0.06 & \llap{0}.10 		      &2.0	       &\llap{Z}O02\\
\llap{1}3& \llap{M}2  & \llap{0}.26  & 0.02     &\llap{0}.05	& 0.21    & 0.01 &\llap{0}.06	      &3.2	       &\llap{Z}O02\\
\llap{1}4& \llap{M}4  & \llap{0}.32  & 0.02     &\llap{0}.05	& 0.51    &  0.17 & \llap{0}.06 		      &\llap{4}6.5     &\llap{Z}O02\\
\llap{1}5& \llap{M}5  & \llap{0}.21  &\llap{$-$}0.14 &\llap{0}.05 & 0.37    &\llap{$-$}0.02 & \llap{0}.06	      &7.8	       &\llap{Z}O02\\
\llap{1}6& \llap{M}4  & \llap{0}.24  &\llap{$-$}0.06 &\llap{0}.05 & 0.35    &  0.01     & \llap{0}.11	      &9.6	       &\llap{Z}O02\\
\llap{1}7& \llap{M}5.5& \llap{0}.33  &\llap{$-$}0.04 &\llap{0}.05 & 0.63    &  0.21 &\llap{0}.05 		      &\llap{6}0.0     &\llap{Z}O02\\
\llap{1}8& \llap{M}6$\dag$ & \llap{0}.40  &\llap{$-$}0.01 &\llap{0}.07 & 0.72	&  0.28 &\llap{0}.08		      & 	       &\\
\llap{1}9& \llap{M}3.5& \llap{0}.30  &0.01      &\llap{0}.05 & 0.54    &  0.22 &\llap{0}.06		      &\llap{2}1.2     &\llap{Z}O02\\
\llap{2}0& \llap{M}3  & \llap{0}.21  &\llap{$-$}0.06 &\llap{0}.05  & 0.16    &\llap{$-$}0.12 &\llap{0}.06	       &4.0		&\llap{Z}O02\\
\llap{2}1& \llap{M}2  & \llap{0}.26  &0.02      &\llap{0}.05  & 0.27    &  0.06 & \llap{0}.06		       &4.1		&\llap{Z}O02\\
\llap{2}2& \llap{M}3  & \llap{0}.19  &\llap{$-$}0.09 &\llap{0}.05  & 0.51    &  0.22 & \llap{0}.07		       &3.6		&\llap{Z}O02\\
\llap{2}3& \llap{M}5$\dag$ & \llap{0}.35  &0.01 &\llap{0}.07	 & 0.52    &  0.12 & \llap{0}.07		       &		&\\
\llap{2}4& \llap{M}4$\dag$ & \llap{0}.26  &\llap{$-$}0.04 &\llap{0}.07 & 0.30	&\llap{$-$}0.04 	& \llap{0}.06	    &	       &\\
\llap{2}5& \llap{M}3.5& \llap{0}.31  &0.02      &\llap{0}.05 & 0.52    &  0.21 &\llap{0}.10		      &4.2	       &\llap{Z}O02\\
\llap{2}6& \llap{M}4$\dag$ & \llap{0}.30  &\llap{$-$}0.01 &\llap{0}.07 & 0.33	&\llap{$-$}0.01        & \llap{0}.08	   &	       &\\
\llap{2}7& \llap{M}5$\dag$ & \llap{0}.30  &\llap{$-$}0.05 &\llap{0}.07 & 0.59	&  0.20        &\llap{0}.11	      & 	       &\\
\llap{2}8& \llap{M}4$\dag$ & \llap{0}.27  &\llap{$-$}0.03 &\llap{0}.07 & 0.77	&  0.43 &\llap{0}.12		      & 	       &\\
\hline										      
\end{tabular}
{$\dag$} objects with spectral types estimated from their colours.\\
\end{minipage}
\end{table}

PMS objects are young and have lower surface gravities than the field dwarfs used to calibrate the empirical relations used in the previous section \citep{luhman99}. We investigated if theoretical models (see Sect.\,2.2) could be used to predict K- and L$'$-band magnitudes for our sample. If, in the $IJKL'$ diagram (Fig.\,\ref{colour_colour}), we compare the colours of objects with no apparent excess with the theoretical isochrones of \citet{baraffe98}, the model $(K-L')$ colours seem to be too blue (even bluer than corresponding giant colours), especially for earlier spectral types ($(I-J) < 1.7$\,mag). In terms of surface gravity, we would expect PMS stars to have colours between main-sequence and giant stars, therefore we opt to use empirical spectral type calibrations to compute $(K-L')$ excesses. Furthermore, a disk frequency based on the comparison with model colours would not be readily comparable to other surveys that tend to use the main-sequence calibration. If model colours were used, the $IJKL'$ diagrams implies that a larger fraction ($\sim$\,80\%) of objects would have an $(K-L')$ excess. 

\begin{figure}
\includegraphics[height=16cm]{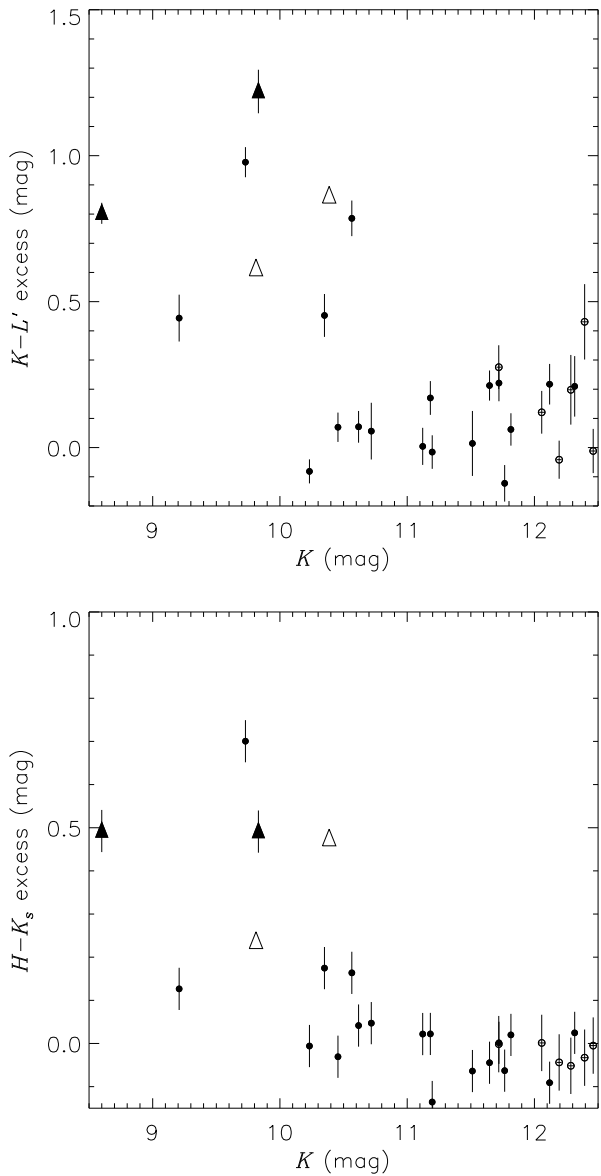}
\caption{$(K-L')$ and $(H-K_{\rm s})$ excesses. These excesses were computed with respect to the targets' predicted colours from their spectral types. The filled circles are objects with known spectral types \citep{osorio02} whereas open circles represent objects with spectral types estimated from their $(I-J)$ and $(H-K_{\rm s})$ colours. The triangles represent the IRAS objects: filled for objects with known spectral type and open for lower limits to the excess (see text). The errors are the uncertainties in the measured and intrinsic $(K-L')$ combined in quadrature (see text); the error bars in $K$ are typically the size of the plotting symbol.}
\label{excess_spt}
\end{figure} 

Spectral types are known for 18 objects in the sample \citep[][ targets labelled ZO02 in Tables\,\ref{obs_table}\,\&\,\ref{excess_table}] {osorio02}. Spectral types are also known for 2 of the IRAS sources. For the 6 objects identified by \citet{kenyon03} spectral types are not known. Their $(I-J)$ and $(H-K_{\rm s})$ colours may be used to estimate spectral types, as none of these objects show considerable excess in the K-band. We use tabulated relations between spectral type and colours (again from \citealt{bessell88}) to define linear relations between spectral type and each colour (these relations are only valid for spectral types later than M2). These are then interpolated and averaged to obtain the spectral type for the sources. The results are in column\,2 in Table\,\ref{excess_table} (indicated by $\dag$). We apply the same procedure to the stars with known spectral types in order to test its validity. For the 10 sources with known spectral type later than M2, we obtain the following: for 6/10 stars we compute the correct spectral type within 0.5\,subclass, for 3/10 within 1\,subclass and for 1/10 within 1.5\,subclass. So we are confident that our computed spectral types are mostly correct to within 1\,subclass.

Even though these may not be the most accurate spectral type determinations, they are easily adequate for the purposes of computing $(K-L')$ excesses. Between spectral types M2 and M6, one subclass is equivalent to at most a variation of 0.055\,mag in photospheric $(K-L')$ colour and 0.05\,mag in $(H-K_{\rm s})$. We adopt these values as estimates of the errors in the excesses introduced by this procedure; spectral types from the literature are uncertain by 0.5 subclass. For each target we have now spectral types and we have computed $(H-K_{\rm s})$ and $(K-L')$ excesses that are listed in columns\,4\,\&\,7 in Table\,\ref{excess_table}. For the 2 IRAS sources without known spectral types, we compute lower limits for their excesses by assigning them spectral type M6 (even though their brightness clearly indicates earlier spectral type) and photospheric colours of respectively $(K-L')=0.46$\,mag and $(H-K_{\rm s})=0.40$\,mag.

In Fig.\,\ref{excess_spt} we plot the computed excesses against K-band magnitude for both $(K-L')$ and $(H-K_{\rm s})$. The error bars are obtained by adding in quadrature the photometric uncertainty in the colours with the uncertainty introduced by the spectral type determination. As expected all 4 IRAS objects have very large excesses in the L$'$-band, consistent with their mid-IR colours \citep{oliveira03b}. We consider that an object has a circumstellar disk if it exhibits an excess detected at a 2-$\sigma$ level. Thus 11/24 (or 46\%) objects show a $(K-L')$ excess. If we arbitrarily adjust for the systematic effect discussed on Sect.\,3.2, then the computed excesses would be larger by about 0.04\,mag and 14/24 objects would have a significant excess. 

\section{Discussion}

\subsection{How representative is this sample?}

The sample we discuss here is not complete, but we believe it is representative. As we discuss below, there is no serious bias either for or against the detection of disks in this sample (apart from the 4 IRAS sources).

\subsubsection{X-ray selection bias}

The equivalent width (EW) of the H$\alpha$ emission line has traditionally been used to classify T Tauri stars (TTS): classical T Tauri stars (CTTS) are defined as having large EW[H$\alpha$] as evidence of circumstellar accretion and thus of the presence of a circumstellar disk; weak-lined T Tauri stars (WTTS) have chromospheric EW[H$\alpha$] thus showing no signatures of accretion. CTTS are found to be underluminous in X-rays when compared with WTTS in the same star-forming regions \citep[e.g.\,][]{neuhauser95,flaccomio03}. This would imply that surveys for circumstellar disks in X-ray selected samples may be biased towards objects with no accretion disk signatures, depending on the sensitivity of the X-ray surveys.

But many PMS stars classified as WTTS based on their H$\alpha$ emission were actually found to have circumstellar disks. \citet{haisch01b} found that a large fraction of WTTS in IC\,348 have considerable $(K-L)$ excesses, i.e.\ WTTS are not necessarily naked TTS. Furthermore, \citet{preibisch02} provided statistically significant evidence that accreting stars in IC\,348 have lower X-ray luminosity when compared with non-accreting objects, but found {\it no evidence} for a difference between stars with or without $(K-L)$ excess. This suggests that X-ray selection will not impair the detection of circumstellar disks in the L-band, even though it might be biased against strongly accreting objects. In fact, as we discuss in the next section, about half of the X-ray selected cluster members have large $(K-L')$ excesses, similar to the overall disk fraction.

\subsubsection{Spectroscopic selection bias}

Another potential source of bias has to do with the spectroscopic identifications of cluster members. The presence of the Li\,{\sc i} 6708\AA\ feature in  stellar spectra is a sure sign of youth at least in the mass range we are considering here. However, photospheric spectral lines in CTTS can be heavily veiled at optical wavelengths \citep[e.g.\,][]{gullbring98}. In particular, the hot continuum attributed to disk accretion can fill-in the Li\,{\sc i} 6708\,\AA\ line  \citep*[e.g.\,][]{magazzu92}. The sample was selected on the basis of the detection of the lithium feature, therefore our estimated disk frequencies could be regarded as a lower limit. \citet{kenyon03} found a few objects in their sample of cluster candidates that show no lithium in their spectra but have radial velocities and Na\,{\sc i} doublet strengths consistent with cluster membership. However, their measured H$\alpha$ line widths are not consistent with accreting PMS objects \citep{white03}. No evidence has been found for heavily-veiled strongly-accreting cluster members, therefore we consider that the spectroscopic selection of cluster members is not a significant source of bias.

\begin{figure*}
\includegraphics[height=8cm]{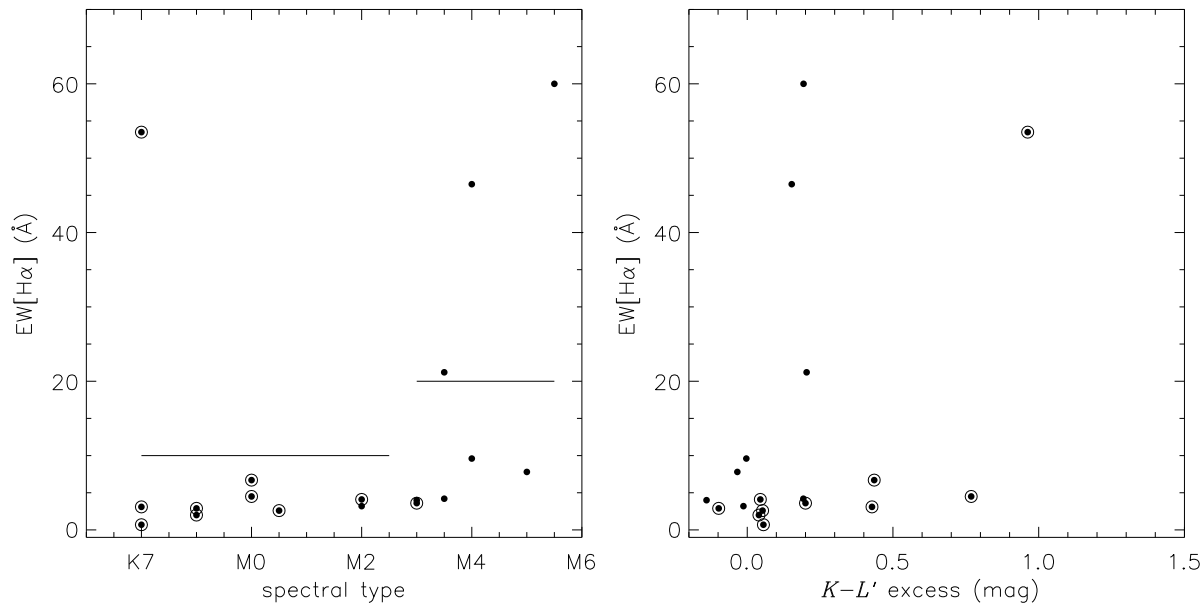}
\caption{EW[H$\alpha$] against spectral type and $(K-L')$ excess. We only have H$\alpha$ measurements for stars from \citet{osorio02}. Double-circled objects were originally identified as ROSAT X-ray sources. The solid lines represent the levels of chromospheric activity for a range of spectral types (see text).}
\label{halpha}
\end{figure*} 

\subsection{L$'$-band excess versus other disk indicators}

In this section we compare $(K-L')$ excesses with other disk indicators. \citet{osorio02} describe spectra of 18 of the stars in our sample (Table\,\ref{obs_table}); in particular they have measured equivalent widths for the H$\alpha$ emission and several forbidden emission lines. A classification criterion for PMS stars based on EW[H$\alpha$] has to take into account the different levels of chromospheric H$\alpha$ emission observed for different spectral types \citep{martin98}. \citet{white03} have proposed the following (empirical and spectral-type dependent) classification: a PMS star is classified as a CTTS if EW[H$\alpha$]$>$3\,\AA\ for K0$-$K5 stars, EW[H$\alpha$]$>$10\,\AA\ for K7$-$M2.5 stars, EW[H$\alpha$]$>$20\,\AA\ for M3$-$M5.5 stars and EW[H$\alpha$]$>$40\,\AA\ for M6$-$M7.5 stars. This classification was devised using high-resolution spectra, so when applying it to intermediate- and low-resolution spectra (as is the case here) we have to take into account that EW[H$\alpha$] is probably overestimated due to blending with a nearby TiO feature. The presence of forbidden lines (e.g.\, [O\,{\sc i}], [N\,{\sc ii}] and [S\,{\sc ii}]) in the spectrum of a PMS star is evidence of jets and outflows that are also related with accretion processes and circumstellar disks \citep[e.g.\,][]{edwards87}.

In Fig.\,\ref{halpha} (left), we plot EW[H$\alpha$] and spectral types for the objects from \citet{osorio02}; on the right we plot EW[H$\alpha$] against $(K-L')$ excesses. We can see that 4 objects show EW[H$\alpha$] in excess of what is expected for their spectral types; these 4 objects have $(K-L')$ excesses of the order of 0.17$-$0.98\,mag, consistent with the presence of circumstellar disks (objects 3, 14, 17 and 19 in Tables\,\ref{obs_table} \& \ref{excess_table}). Of these 4 objects, 3 exhibit spectra with forbidden line emission (objects 3, 14 and 19). Of the 10 objects that were X-ray selected, only one object has a large EW[H$\alpha$]; half of these objects have a $(K-L')$ excess. The IRAS sources have, as expected, large $(K-L')$ excesses, as they also have large mid-IR excesses \citep{oliveira03b}. 

There are also 3 objects that have $(K-L')$ excesses larger than 0.4\,mag, but do not show large EW[H$\alpha$]. This again illustrates that $(K-L')$ is the most efficient and robust disk indicator. H$\alpha$ emission is known to be extremely variable \citep{guenther97} and objects that have $(K-L')$ excesses indicative of circumstellar disks could alternate from episodes of high accretion activity (CTTS state) to episodes of undetected accretion activity (WTTS). More interestingly, if we interpret this as an evolutionary sequence (i.e.\ the disk-star system evolves from an accreting system, through a weakly or non-accreting system, to a naked PMS star), this could be an indication that circumstellar disks can survive for some time after accretion onto the stellar surface has stopped. The $\sigma$\,Ori cluster is at a crucial stage of disk destruction (see next section) and we would indeed expect to find a mixture of objects that are still accreting and others where accretion has strongly diminished or ceased. 

\subsection{Comparison with other young clusters}

The most complete analysis of clusters' disk frequencies is described in \citet[][ and references therein]{haisch01}. They compile the results of several cluster surveys in the K- and L-bands. These authors have consistently used the same method to determine which objects possess a $(K-L)$ excess: they place cluster members in a $JHKL$ colour-magnitude diagram --- only those objects with $K$ magnitudes brighter than the completeness limit of the $L$ survey are considered, to assure that stellar photospheres are detected in both bands; they count the stars that lie to the right of the reddening vector that passes through the position in this diagram of an M5 main-sequence star. We can apply exactly the same procedure in Fig.\,\ref{colour_colour} (top right): 13/24 stars (54\,$\pm$\,15)\%\footnote{Statistical errors are conservatively estimated as $\sqrt{N_{\rm disk}}/N$.} are to the right of the reddening band. This result depends on the adopted reddening law --- the reddening law defines the position of the boundary between objects with and without excesses --- but the different determinations are  within the quoted statistical errors \citep{haisch01}. Depending on the spectral type distribution of the sample, this method might underestimate the true disk frequency particularly for earlier spectral types \citep{lada00}.

According to this technique, the $\sigma$\,Ori cluster has a (54\,$\pm$\,15)\% $JHKL'$-excess disk frequency, at an age of 3$-$8\,Myr. How does this compare with other clusters? Such comparison can be done by placing the $\sigma$\,Ori cluster in Fig.\,1 from \citet{haisch01} that plots the fraction of $JHKL$ excess objects against stellar age for several young clusters and associations. In terms of its age, the $\sigma$\,Ori cluster can be compared with NGC\,2264 (age\,$\sim$\,3.2\,Myr, disk frequency 52\%\,$\pm$\,4\%) and NGC\,2362 (age\,$\sim$\,5\,Myr, disk frequency 12\%\,$\pm$\,10\%). Crucially, the measurements for these two clusters --- particularly NGC\,2362 --- allowed the authors to estimate that the overall disk lifetime is about 6\,Myr, with 50\% of the disks dispersed by about 3\,Myr. In this scenario, the $\sigma$\,Ori cluster can play a key role to better constrain disk destruction timescales. In \citet{haisch01} analysis, the ages of the clusters were determined using different PMS models and these authors estimate that this introduces an overall systematic uncertainty in the ages of the order of 1.2\,Myr. The age of $\sigma$\,Ori is also not well established (see discussion Sect.\,2.2): 3$-$5\,Myr is favoured by several authors (\citealt{bejar99,oliveira02,jayawardhana03}; this work) with an upper limit of 8\,Myr \citep{osorio02}. If the age of the cluster is $\la$\,4\,Myr then our estimate of disk frequency is consistent with the above mentioned timescales. However, if the cluster is older than 4\,Myr, then our result suggests a slower disk destruction, i.e.\, a longer overall disk lifetime. The NGC\,2264 and NGC\,2362 surveys are for objects more massive than $\sim$\,0.85\,M$_{\sun}$ (i.e.\ earlier spectral types) while our sample populates the range 0.13$-$1.0\,M$_{\sun}$. As mentioned above, this technique might be at fault for earlier spectral types. Furthermore, if disk-destruction timescales are mass-dependent then one has to be careful when comparing these results. 

\citet{lyo03} analysed $(K-L)$ excesses in the $\sim$\,9\,Myr, sparsely-populated $\eta$\,Chamaeleontis cluster. They found that of the 12 late-type stars in the central part of the cluster 7 objects have $(K-L)$ excess. Their mass range is comparable to the mass range of our sample, but even so the disk frequency in $\eta$\,Cha is remarkably high for its age. Environmental effects might explain such result: disk destruction timescales might be controlled also by the stellar density in the star-forming region and by the photoevaporation power of nearby O-stars. If this is the case, then $\eta$\,Cha cannot be straight forwardly compared with the other clusters.

\subsection{Comparison with disk model predictions}

\citet{wood02} investigates the observational signatures of circumstellar disks in a simple evolutionary scenario by which disk mass (of small particles) decreases with time (for a stellar effective temperature of 4000\,K, spectral type K7-M0). Except for high disk masses, the main contribution to the spectral energy distribution (SED) is from reprocessing of starlight (or disk irradiation), as inferred from observations \citep{hartmann98}. For the $(K-L)$ excess (or $\Delta(K-L)$) their circumstellar disk model predicts that: {\it i}) large excesses ($\Delta(K-L) \ga$ 0.7\,mag) can only be achieved with the contribution of accretion luminosity from massive disks; {\it ii}) for passive disks, $\Delta(K-L)$ is insensitive to disk mass over several orders of magnitude; {\it iii}) for disk masses lower than 10$^{-7}$\,M$_{\sun}$, $\Delta(K-L)$ decreases rapidly. 

In our sample, 9 of the objects have spectral types K7$-$M0.5 and 3 of these objects have large excesses ($\ga$ 0.7\,mag) that can hint at a contribution from accretion luminosity to the SEDs. Only one of these objects shows signatures of accretion in its spectrum (Sect.\,5.2) but variability likely plays an important role. The objects with later spectral types tend to have modest excesses ($\la$ 0.4\,mag), consistent with disk irradiation.  

\section{Summary}

In this paper we present new K- and L$'$-band photometry for a representative sample of $\sigma$\,Orionis cluster members. Different methods and disk-excess criteria will provide somewhat different disk frequencies. We computed a disk frequency for this cluster of (54$\pm$15)\%, if measured directly from the $JHKL'$ colour-colour diagram, or (46$\pm$14)\%, if excesses are computed with respect to predicted photospheric colours (according to the objects spectral types, 2-$\sigma$ excess detections). Therefore, when trying to determine a disk destruction timescale, one has to be consistent in the methods used.

An overall disk lifetime of 6\,Myr has been proposed, but this relies heavily on the age and disk frequency derived from NGC\,2362 (age\,$\sim$\,5\,Myr and disk-frequency\,$\sim$\,12\%, \citealt{haisch01}). The disk frequency for the $\sigma$\,Ori cluster could increase support for this timescale, but we find that the crucial factor is not so much the derived disk frequency, but the {\it age determination} of the cluster. As is the case for most young clusters \citep{hartmann01}, the age of the $\sigma$\,Ori cluster is uncertain, with a likely age in the range 3$-$5\,Myr \citep{oliveira02,osorio02,jayawardhana03}. If the cluster age is about 3\,Myr then it lends support to the above mentioned disk lifetimes. However, if the cluster is older than 4\,Myr then it points towards slower disk destruction timescales. 

If disk dissipation timescales are mass dependent then the disk frequencies derived for the different clusters are not so readily comparable. Another factor that is probably relevant is the environment of the star forming region, in particular the rates of close stellar encounters and the amount of ionizing radiation produced by massive (O-type) cluster members.

\section*{Acknowledgments}
We thank the staff of the United Kingdom Infrared Telescope (UKIRT) for their support during the observing run. The UKIRT is operated by the Joint Astronomy Centre  on behalf of the U.K. Particle Physics and Astronomy Research Council (PPARC). This publication makes use of data products from the Two Micron All Sky Survey, which is a joint project of the University of Massachusetts and the Infrared Processing and Analysis Center/California Institute of Technology, funded by 
the National Aeronautics and Space Administration and the National Science 
Foundation. JMO acknowledges the financial support of PPARC.

\bsp

\label{lastpage}

\end{document}